# Skeleton Detection Using Dual Radars with Integration of Dual-View CNN Models and mmPose

Masaharu Kodama

Graduated School of Computer and Information Sciences,

Hosei University, Tokyo, Japan

masaharu.kodama.6j@stu.hosei.ac.jp

Runhe Huang

Faculty of Computer and Information Sciences,

Hosei University, Tokyo, Japan

rhuang@hosei.ac.jp

*Abstract*— Skeleton detection is a technique that can be applied to a variety of situations. It is especially critical identifying and tracking the movements of the elderly, especially in real-time fall detection. While conventional image processing methods exist, there's a growing preference for utilizing point clouds data collected by mmWave radars from viewpoint of privacy protection, offering a non-intrusive approach to elevate safety and care for the elderly. Dealing with point cloud data necessitates addressing three critical considerations. Firstly, the inherent nature of point clouds—-rotation invariance, translation invariance, and locality—-is managed through the fusion of PointNet and mmPose. PointNet ensures rotational and translational invariance, while mmPose addresses locality. Secondly, the limited points per frame from radar require data integration from two radars to enhance skeletal detection. Lastly, inputting point cloud data into the learning model involves utilizing features like coordinates, velocity, and signal-to-noise ratio (SNR) per radar point to mitigate sparsity issues and reduce computational load. This research proposes three Dual View CNN models, combining PointNet and mmPose, employing two mmWave radars, with performance comparisons in terms of Mean Absolute Error (MAE). While the proposed model shows suboptimal results for random walking, it excels in the arm swing case.

## I. Introduction

Research on skeleton detection commonly uses cameras and LiDAR due to their high accuracy and the wealth of information each frame provides. However, these devices have drawbacks. Camera images contain privacy-sensitive information such as fingerprints and faces. LiDAR, which uses visible light for point cloud data, can be obstructed in rainy, foggy, or smoky conditions, resulting in poor-quality data [1]. Cameras also suffer in inclement weather. Moreover, LiDAR struggles with high-reflectivity objects like white surfaces, as seen in the inability of Uber's self-driving cars in Arizona to detect white trailers [2]. I propose using millimeter-wave radar for skeleton detection. Millimeter-wave radar preserves privacy and is unaffected by rain, fog, or smoke. It can also penetrate low-reflectivity objects. However, its point cloud data is coarser than that of cameras or LiDAR, requiring careful consideration.

This study proposes a CNN model for skeleton detection in posture estimation systems for the elderly and patients. In Japan, the aging population has led to a shortage of caregivers [3]. Monitoring systems using AI and devices are crucial for alleviating this burden. However, wearable devices may be removed, and cameras may be rejected due to privacy concerns. Thus, millimeter-wave radar, ensuring non-contact and privacy protection, is significant. Millimeter-wave radar can detect through fabric, reducing concerns about being watched. It is also gaining attention as on-board radar for autonomous vehicles. High-precision skeleton detection with millimeter-wave radar may enable safer autonomous driving. Therefore, this approach holds promise for various systems and is a valuable research area.

The remainder of this paper is organized as follows: Section II introduces literature review. Section III explains the method of generating point cloud from radar data. Section IV presents Issues to be aware of when handling point cloud data. Section V proposes the solution to problems of skeleton detection using point cloud data. Section VI reviews experiments and results. Section VII discusses evaluation and investigation of results. Finally, Section VIII concludes the paper with a summary of our findings and suggestions for future research directions.

## II. Literature Review

Various studies on skeleton detection using radar have been conducted. In 2015, MIT researchers introduced RF-Capture [4], a system that detects coarse skeletons through walls using RF signals and tracks the 3D positions of keypoints. While it achieved user identification through walls, it couldn't achieve complete skeleton detection over time. In 2018, MIT researchers proposed RF-Pose [5], which takes 2D heatmaps of RF signals from vertical and horizontal directions as input, processed through an encoder and decoder. The training data consisted of keypoint estimations based on RGB data captured by a camera. RF-Pose achieved 93.3% similarity at a threshold of 0.5 in OKS, compared to the baseline model OpenPose [6] at 77.8%. Based on RF-Pose, A. Sengupta et al. proposed mmPose [7], which used millimeter-wave radar. The input to mmPose is 16×16 RGB image data from two planes, with coordinates represented by the R and G values, and signal intensity as the B value. mmPose estimates 3D keypoint coordinates, evaluated with MAE, showing localization accuracy of 3.2cm (Depth), 2.7cm (Elevation), and 7.5cm (Azimuth). Compared to RF-Pose, mmPose showed improved accuracy in Depth and Elevation. Subsequently, A. Sengupta et al. proposed mmPose-NLP [8], based on the Seq2Seq model, incorporating temporal data learning. This model achieved MAE of 2.14cm (Depth), 1.49cm (Elevation), and 1.76cm (Azimuth). Existing research indicates the use of data from vertical and horizontal planes and models utilizing encoders and decoders for temporal learning. This paper proposes a model using data from two planes as input.

### III. POINT CLOUDS GENERATION

#### A. FMCW Radar and Signal Processing

The millimeter-wave radar used in this study employs Frequency Modulated Continuous Wave (FMCW) radar. This radar method continuously increases the frequency of millimeter waves at regular time intervals, allowing for the measurement of distance, velocity, angle, and other properties of objects. Furthermore, knowing the distance and velocity enables the generation of point cloud coordinate data. In this section, we explain the signal processing mechanism employed to obtain point cloud information (coordinates, velocity) used in this study. The signal processing mechanism described here is based on reference [9][10].

#### B. Methods of Distance Calculation

The FMCW radar consists of a transmitting antenna and a receiving antenna. The millimeter waves transmitted from the transmitting antenna bounce off objects and are received by the receiving antenna. The internal system of the FMCW radar continuously changes the frequency of the signal it transmits linearly from the starting frequency to the ending frequency over a fixed time interval $T_c$. This signal is called a chirp signal. When measuring the distance between an object and the radar, we utilize information from both the chirp signal transmitted from the transmitting antenna (TX chirp) and the chirp signal received by the receiving antenna (RX chirp). τ represents the time delay, which is proportional to the distance between the object and the radar and can be expressed as shown in Equation (1).

$$\tau = \frac{2d}{c} \quad (1)$$

where, $d$ is the distance between the object and the radar, and $c$ is the speed of light. We derive something called the Intermediate Frequency (IF) signal from the difference between the TX chirp and RX chirp over the interval between $\tau$ and the fixed time interval $T_c$. The expression for the IF signal is a sinusoidal wave, as shown in Equation (2).

$$A sin(2\pi f_0 t + \phi_0) \quad (2)$$

In this case,

$$\phi_0 = 2\pi f_c t \quad (3)$$

By the relationship between Equation (1) and Equation (3), we can derive Equation (4).

$$\phi_0 = \frac{4\pi d}{\lambda} \quad (4)$$

Furthermore,

$$f_0 = \frac{S2d}{c} \quad (5)$$

where, $S$ is the frequency of the IF signal. Although the sinusoidal wave of the IF signal is received multiple times for a certain frame, the synthesized waveform of these constitutes the IF signal for that frame. By performing Fourier transform on this synthesized waveform, we can obtain frequency spectra with unique peaks for each received wave, corresponding to the reflected points' distances. For instance, if the result of Fourier transform on the IF signal yields frequency spectra $f_1, f_2, f_3$, the distances $R_1, R_2, R_3$ of the points where each wave is reflected can be calculated as shown in Equation (6).

$$R_1 = \frac{f_1}{2c}, R_2 = \frac{f_2}{2c}, R_3 = \frac{f_3}{2c} \quad (6)$$

#### C. Methods of Velocity Calculation

FMCW radar transmits modulated waves at intervals of $T_c$. The velocity can be determined by utilizing the phase difference between the received waves RX1 and RX2 at the time of transmitting wave TX1 and the transmitting wave TX2 at a time $T_c$ later. When there is a time shift of $\Delta\tau$ in receiving RX1 and RX2 due to the movement of the object, the phase difference of the Intermediate Frequency (IF) signal is as shown in Equation (7).

$$\Delta\phi = 2\pi f_c \Delta\tau \quad (7)$$

Furthermore, Equation (8) can be derived from Equation (1) and Equation (5) as follows:

$$\Delta\phi = \frac{4\pi v T_c}{\lambda} \quad (8)$$

From Equation (8), the velocity v can be derived as shown in Equation (9).

$$v = \frac{\lambda \Delta\phi}{4\pi T_c} \quad (9)$$

#### D. Methods of Angle Calculation

Angles can be measured using the phase difference of each receiving antenna. If the distance between RX1 and RX2 is $l$, the distance between RX2 and the object is extended by $l sin\theta$. Assuming that the distance between the radar and the object is sufficiently large, the difference in distance and Equation (4) leads to:

$$\Delta\phi = \frac{2\pi l sin\theta}{\lambda} \quad (10)$$

From this, the angle can be derived as follows:

$$\theta = sin^{-1}\left(\frac{\lambda \Delta\phi}{2\pi d}\right) \quad (11)$$

#### E. Point Clouds Data

Point cloud data can be generated using distances, angles, and velocities calculated as described in sections B, C, and D. Knowing the distance from the radar and the angle relative to the radar provides coordinates in 3D space. Additionally, velocities obtained are represented by varying the intensity of color on the points. In this study, the acquisition and generation of point cloud data were conducted using Python and the radar's API.

When handling point cloud data in machine learning, three properties need to be considered. Firstly, there is the property of invariance to order. Invariance to order means that the order of inputting point cloud data into the model does not affect the learning outcome. For example, if point cloud data is inputted in the order of increasing X-coordinate values, the learning process may inadvertently rely on that specific order. Secondly, there is the property of translation invariance. Translation invariance means that the output remains unchanged even when point cloud data undergoes parallel or rotational translations. Thirdly, there is the property of locality. Locality implies that points close to each other have strong correlations, while points farther apart have weaker correlations.

## IV. PROBLEMS OF USING POINT CLOUD DATA

When performing attitude estimation using millimeter-wave radar, there are three points that must be considered. Firstly, the three properties inherent in point cloud data, namely invariance to order, translation, and locality. Secondly, the limited number of point clouds obtained from a single irradiation of millimeter waves. Thirdly, when conducting machine learning on three-dimensional data by dividing the bounding box into cells, there is a risk of sparsity in the data and unnecessary computational overhead.

### A. Data Density and Computational Complexity

When dealing with 3D point cloud data, one approach is to provide a bounding box and divide it into cells. In this case, if the number of points obtained is low, the data may become sparse, leading to wastage of computational resources. Moreover, sparsity in data can result in gradients becoming close to zero during machine learning, posing a problem. Additionally, excessive computational overhead can make real-time skeleton detection challenging.

### B. Sparsity of Radar Data

Millimeter-wave radar presents a challenge in that it generates a limited number of point clouds upon a single irradiation of millimeter waves. While this limitation may not be problematic for object tracking, it becomes critical for skeleton detection, where a larger amount of point cloud data is required. This necessity arises because achieving high accuracy demands capturing both local and global variations in the point cloud data. For instance, predicting the movements of body extremities such as arms and legs requires local point cloud information, whereas forecasting overall body movements resulting from walking necessitates global point cloud information.

## V. PROPOSED SOLUTIONS TO THE PROBLEMS

In this section, we will discuss a proposed approach that addresses the three issues raised in Section IV. As a solution, we propose two approaches: the integration of PointNet and mmPose and the use of two radars.

### A. Existing PointNet [11] and mmPose [12]

PointNet is a network primarily utilized for classification and segmentation of point cloud data. This network is designed taking into consideration two of the three properties of point cloud data mentioned in III-E: invariance to order and invariance to translation. One of the significant features of this network is the presence of MaxPooling layers. MaxPooling, by outputting the maximum value of the input, exhibits a property that is invariant to the order of the inputs, thus addressing invariance to order. Additionally, another crucial component is the presence of Transform-Net. This network applies affine transformations to the input point cloud to approximately consider translation invariance. The overall structure of PointNet incorporates multiple uses of MLP (Multi-Layer Perceptron) and MaxPooling.

mmPose, introduced by A. Sengupta et al. in 2020, is a network that predicts and generates skeletal data in real-time from point clouds obtained from millimeter-wave radar. This network possesses a significant feature: the utilization of two image data sets. It voxelizes the 3D point cloud data and converts it into image data viewed from two planes. This enables a substantial reduction in computational complexity compared to directly using voxel data for training. The network architecture employs CNNs of the same size to maintain the information content of the image data while training.

### B. Integration of PointNet and mmPose

As a proposed model, we combine PointNet and mmPose. The proposed model's layer structure is depicted in Figure 1. The input to the model consists of point cloud information viewed from the $xy$ and $yz$ directions, each with a size-4 feature containing coordinates, Doppler velocity, and Signal-Noise Ratio (SNR). After inputting, we perform affine transformation using TNet and pass it through CNNs. Then, we integrate the output of CNNs for each plane and flatten it. Finally, we pass the flattened data through MLP for output. I aim to predict up to 32 joints, with each having $x$, $y$, and $z$ coordinates, resulting in a maximum of 96 output labels. The ground truth data, collected using Azure Kinect DK, comprises coordinates of joints, and we train the model to minimize the Mean Squared Error (MSE) between the ground truth and output. I normalized the ground truth data during gradient computation. The input size is determined by the maximum number of points in point cloud data multiplied by the number of features. While mmPose uses 16×16 image data as input, we maintain consistency with PointNet's format to preserve its properties. In a typical PointNet, the features consist of three-dimensional coordinates. However, in this study, we introduced velocity and SNR as additional features. By incorporating these two features, we aim to improve the accuracy of skeleton detection. Introducing velocity allows the model to learn considering the speed of each point, while introducing SNR aims to exclude points that are likely noise. Since normalizing SNR slightly improved the results, we performed normalization on it.

### C. The Use of Two Radars

To address the issue of sparse point clouds from millimeter-wave radar, we employed two millimeter-wave radars. One radar was positioned at a height of 1m, tilted downward by 15 degrees, while the other was positioned at a height of 2m, tilted downward by 20 degrees (Figure 2). This setup increases coverage in the vertical direction [13]. The higher radar collects point cloud data focusing on the upper body, while the lower radar focuses on the lower body, facilitating the capture of overall body features.

To create a dataset containing point cloud data collected from two millimeter-wave radars and joint coordinate data obtained from one Azure Kinect DK, synchronization was essential. Since both millimeter-wave radars were operated on a single PC, we set a frame time interval and integrated the data based on timestamps. Azure Kinect DK requires a GPU, thus necessitating a separate PC. Therefore, it was crucial to synchronize the time between the PC controlling the millimeter-wave radars and the

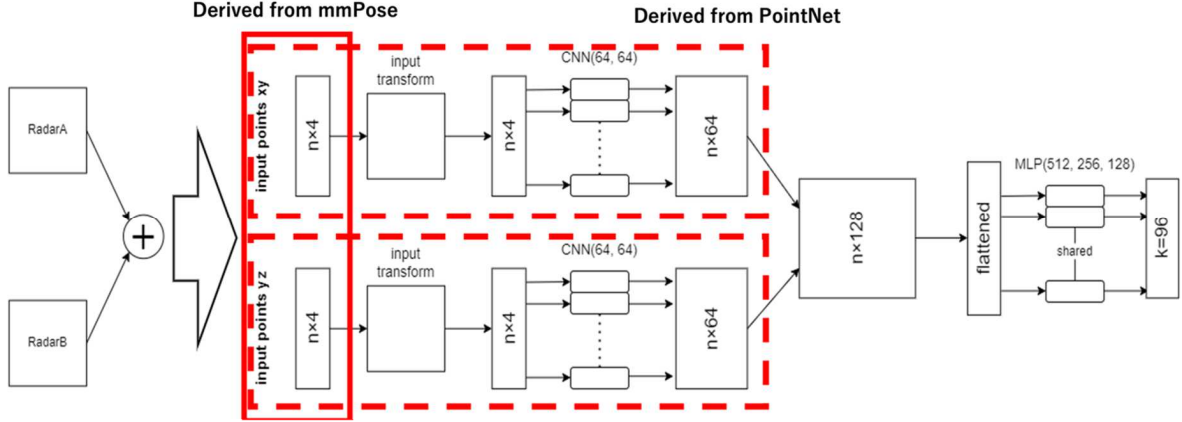

Fig. 1. Dual View CNN Model Based on PointNet and mmPose Using Two Radars

PC operating Azure Kinect DK. I synchronized both PCs' times using the same time server and saved the timestamps for data integration. Subsequently, we conducted data integration based on these timestamps. The integrated point cloud data underwent clustering using DBSCAN (Density-Based Spatial Clustering of Applications with Noise). Clustering helped remove noise from the data. Although millimeter-wave radars utilize the Constant False Alarm Rate (CFAR) algorithm as a standard feature to remove noise during point cloud acquisition, residual noise may still be present, necessitating clustering using DBSCAN. DBSCAN is a density-based clustering method. The parameters to be set include the minimum number of points to be considered as a cluster and the distance (radius) within which any two points are considered to be in the same cluster.

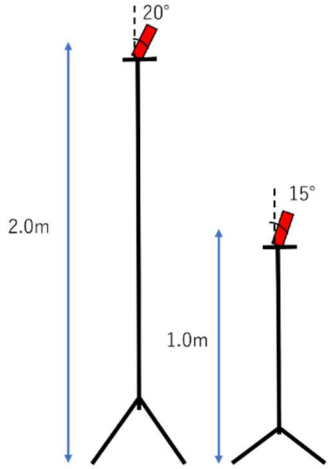

Fig. 2. FMCW radar placement

## VI. EXPERIMENTS AND RESULTS

### A. Experiment Environment

The experiments were conducted indoors, where each of the two subjects performed four actions randomly: "walking towards the radar," "walking away from the radar," "swinging the right arm," and "swinging the left arm." The radar was positioned as described in Section VI-B, and the subjects moved along a straight line at distances ranging from 1.9 to 3.5 meters from the radar. The Azure Kinect DK was positioned at a height of 1.1 meters, aligned with the millimeter-wave radar on the same straight line. For each subject, a total of 5,500 frames were prepared for the training/validation dataset, and a total of 1,700 frames were prepared for the testing dataset. The training/validation dataset was split into an 8:2 ratio for usage.

### B. Experiment Methods

During the experiments, it was observed that the Mean Absolute Error (MAE) between the predicted keypoints and the ground truth keypoints was notably higher for certain keypoints, specifically the wrist, palm, and fingertips. This discrepancy was attributed to the sparse point clouds obtained from reflections on the hands, resulting in larger MAE values for hand keypoints. To address this issue, training was conducted using 22 keypoints excluding those corresponding to the hand, and the results were compared with other three models. The first experiment involves inputting point cloud data collected using two radars into the proposed model's CNN section, which is modified to MLP, as shown in Figure 1, for training. This experiment aims to verify the effectiveness of using CNN by comparing it with the proposed model. This experiment is referred to as **Experiment 1**. The second experiment involves inputting point cloud data collected using two radars into a pure PointNet without a dual-view structure for training. This experiment aims to validate the effectiveness of the dual-view structure by comparing it with the proposed model. This experiment is referred to as **Experiment 2**. The third experiment involves inputting point cloud data collected using one radar into the learning network of the proposed model for training. This experiment aims to validate the effect of the number of radars by comparing it with the proposed model. This experiment is referred to as **Experiment 3**.

The evaluation metrics consist of the overall mean average precision, the mean average precision of the arm portion in frames where arms are swung, the mean average precision of the lower body, and the mean average precision in the Depth direction. Here, since walking direction is only in the anterior-posterior direction, Depth represents the distance from the radar to the body. The overall and Depth-direction mean average precisions evaluate the global prediction accuracy of the entire body, while the mean average precisions of the arm portion in frames where arms are swung and the lower body evaluate the prediction accuracy of local parts. The mean average precisions for each case are shown in TABLE Ⅰ. Additionally, Figure 3 depicts the visualization of keypoints for proposed method. Although we attempted a method to extract 3D bounding boxes, it was excluded from evaluation due to frequent occurrences of gradient disappearance, rendering it incomparable.

## VII. EVALUATION AND INVESTIGATION OF RESULTS

### A. CNN vs. MLP

When comparing the proposed method with the Experiment 1, it is observed that the mean average precision of all 22 keypoints, the mean average precision of the lower body, and the mean average precision in the Depth direction are better for this experiment. However, the mean average precision when swinging arms is better for proposed method. Therefore, it can be said that experiment excels at detecting global body movements, while proposed method is better at detecting local movements. Indeed, as shown in Figure 3 and TABLE Ⅱ, the action of swinging arms can be detected. Figure 3 shows the outputs for all of the hand waving movements. Although there are differences in the degree to which the arms are raised, it is clear that they are able to recognize that the arms are being raised. It was found that Experiment 1 failed to detect arm swinging when visualized. This result can be attributed to the suitability of CNN for extracting local features.

### B. Single View vs. Dual View

When comparing the proposed method with Experiment 2 that takes 3D coordinates as input, it is observed that proposed method performs better in terms of accuracy when swinging arms and the lower body. However, this experiment achieves better overall and Depth-direction accuracy. Therefore, it can be said that the proposed method is better at detecting keypoints requiring local features from point cloud data, while this experiment is more adept at detecting global features of the body. However, in the case of Experiment 1, where the layers are MLP instead of CNN, except when swinging arms, that achieves higher accuracy than Experiment 2. This suggests that in a dual-view setup, CNN is more effective at capturing local features, while MLP is better at capturing global features.

### C. One Radar vs. Two Radars

When comparing the use of Experiment 3 versus proposed method, it's evident that the average precision when swinging arms is significantly better with two units. This is likely because a single radar unit may not detect subtle movements of the body. Conversely, the average precision of the lower body is better with a single radar unit. This outcome might be due to the coverage of the lower body being higher with one radar unit, but it may not fully cover the movement of the arms due to the height relationship of millimeter-wave radar. Additionally, there's a possibility of interference or crosstalk when using two millimeter-wave radar units. The point cloud data collected using two radar units exhibit a larger disparity in the Depth direction compared to the data collected using a single unit, suggesting that if this disparity is caused by interference or crosstalk, it could explain the difference in performance. Furthermore, since the movement of the lower body is primarily in the anterior-posterior direction during walking, the influence of Depth direction disparity is considered significant.

### D. Overall investigation

Although the proposed method can follow local motion to some extent, it is far from practical accuracy; since "PointNet does not capture local structure induced by the metric" [14], it is thought that the effect of building a model based on PointNet has prevented the extraction of local features from working well. A possible solution to this problem is to use PointNet++ [15], an improved version of PointNet that can learn local structure. We thought that this problem also could be solved by considering not only spatial features but also time-series features. Another problem is that the accuracy drops drastically when a part of the point cloud data is missing [16]. Another possible cause of inaccuracy is the lack of countermeasures for millimeter-wave interference problems [17].

## VIII. CONCLUSION

In this study, an original skeleton detection model was developed and employed alongside two millimeter-wave radars for skeleton detection. While the local skeleton detection accuracy surpassed that of the comparison models, the overall skeleton detection accuracy fell short. Additionally, the localization accuracy was not exceptionally high. As a future goal, we aim to improve the global skeleton detection accuracy while enhancing the overall localization accuracy of each key point by refining preprocessing techniques and models.

TABLE I. Comparison of MAE for each model

|  | MAE across 22 key points | MAE when swinging arms | MAE of the lower body | MAE in the Depth direction |
|---|---|---|---|---|
| Two-radar + CNN (Our model) | 16.1cm | 17.1cm | 12.7cm | 31.5cm |
| Two-radar + MLP (Experiment 1) | 14.7cm | 17.7cm | 12.4cm | 28.4cm |
| Two-radar + Simple PointNet (Experiment 2) | 15.1cm | 17.4cm | 12.9cm | 28.9cm |
| One-radar (Experiment 3) | 14.9cm | 19.5cm | 9.4cm | 28.3cm |

TABLE II. Percentage of correct answers in recognition of arm swing

| Models | Percentage of correct recognition of the arm swing |
|---|---|
| Two-radar + CNN (Our model) | 9.6% |
| Two-radar + MLP (Experiment 1) | 0.0% |
| Two-radar + Simple PointNet (Experiment 2) | 0.0% |
| One-radar (Experiment 3) | 2.2% |

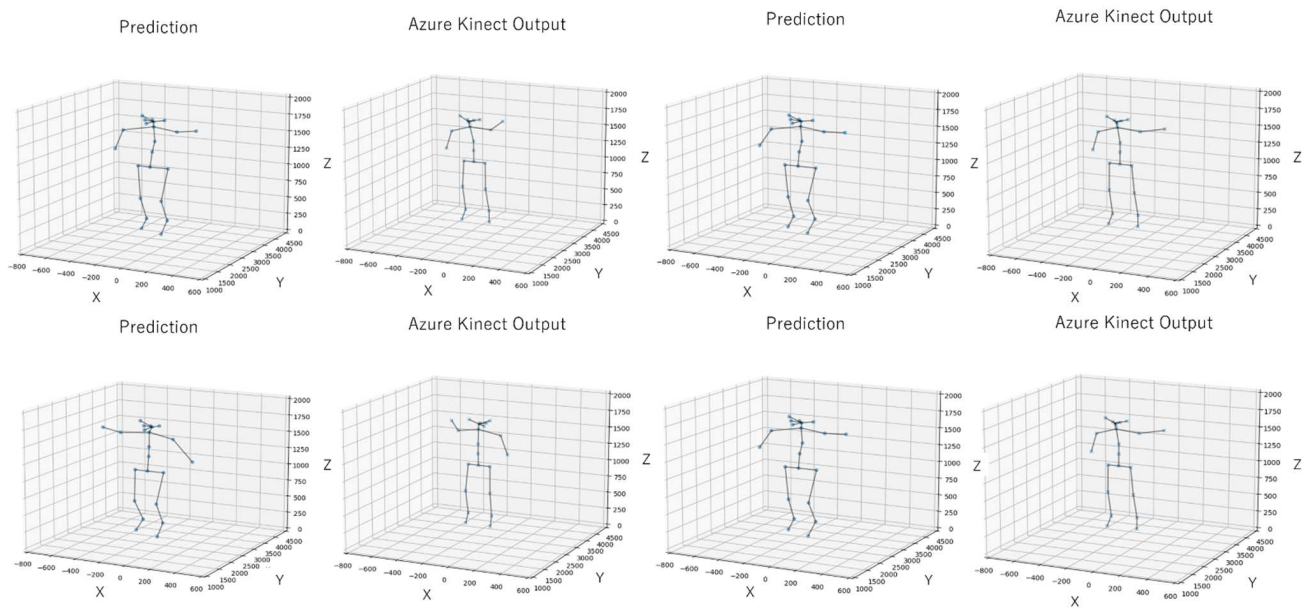

Fig. 3. Comparison of the output of our proposed method with Azure Kinect keypoint coordinates

REFERENCES

[1] S. R. E. Datondji, Y. Dupuis, P. Subirats, and P. Vasseur, "A survey of vision-based traffic monitoring of road intersections,'' IEEE Trans. Intell. Transp. Syst., vol. 17, no. 10, pp. 2681-2698, Oct. 2016.

[2] The Tesla Team. (Jun. 2016). A Tragic Loss. Accessed: January. 22, 2024. [Online]. Available: https://www.tesla.com/blog/tragic-loss

[3] Cabinet Office. [ Aging Population ] Koureika no joukyou (in Japanese). Available: https://www8.cao.go.jp/kourei/whitepaper/w-2021/html/zenbun/s1_2_2.html

[4] F. Adib, C.-Y. Hsu, H. Mao, D. Katabi, and F. Durand, "Capturing the human figure through a wall," ACM Trans. Graph., vol. 34, no. 6, pp. 1–13, Nov. 2015.

[5] M. Zhao et al., "Through-wall human pose estimation using radio signals," in Proc. IEEE/CVF Conf. Comput. Vis. Pattern Recognit., Jun. 2018, pp. 7356–7365.

[6] Z. Cao, T. Simon, S.-E. Wei, and Y. Sheikh. Realtime multiperson 2D pose estimation using part affinity fields. In Proceedings of the IEEE Conference on Computer Vision and Pattern Recognition, CVPR, 2017.

[7] Arindam Sengupta, Feng Jin, Renyuan Zhang, and Siyang Cao. mm-Pose: Real-Time Human Skeletal Pose Estimation Using mmWave Radars and CNNs. IEEE SENSORS JOURNAL, 2020, VOL. 20, NO. 17.

[8] Arindam Sengupta, and Siyang Cao. mmPose-NLP: A Natural Language Processing Approach to Precise Skeletal Pose Estimation Using mmWave Radars. IEEE Transaction on Neural Networks and Learning Systems, 2022, p. 1-12.

[9] The fundamentals of milimeter wave radar sensors, Accessed : 2024/1/22[Online]. Available: https://www.ti.com/jp/lit/wp/jajy058/jajy058.pdf

[10] Principles and Features of Millimeter Wave Radar (in Japanese), Accessed: 2024/1/22 [Online]. Available: https://www.teldevice.co.jp/semiconductor/technical-info/ehf/

[11] Charles R. Qi, Hao Su, Kaichun Mo, and Leonidas J. Guibas. PointNet: Deep Learning on Point Sets for 3D Classification and Segmentation. arXiv preprint arXiv:1612. 00593, 2016.

[12] Arindam Sengupta, Feng Jin, Renyuan Zhang, and Siyang Cao. mm-Pose: Real-Time Human Skeletal Pose Estimation Using mmWave Radars and CNNs. IEEE SENSORS JOURNAL, 2020, VOL. 20, NO. 17..

[13] Arindam Sengupta, and Siyang Cao. mmPose-NLP: A Natural Language Processing Approach to Precise Skeletal Pose Estimation Using mmWave Radars. IEEE Transaction on Neural Networks and Learning Systems, 2022, p. 1-12.

[14] Charles R. Qi, Li Yi, Hao Su, and Leonidas J. Guibas. PointNet++: Deep Hierarchical Feature Learning on Point Sets in a Metric Space. arXiv pre print arXiv:1706. 02413, 2017, lines 10-11 in Section I.

[15] Charles R. Qi, Li Yi, Hao Su, and Leonidas J. Guibas. PointNet++: Deep Hierarchical Feature Learning on Point Sets in a Metric Space. arXiv pre print arXiv:1706. 02413, 2017.

[16] Qi, Z., Yang, J., Chen, X., Li, S., Feng, Z., & Shi, Y. mmBody Benchmark: 3D Body Reconstruction Dataset and Analysis for Millimeter Wave Radar. arXiv preprint arXiv:2209. 05070, 2022, lines 2-3 in Section 5.3.

[17] Li, X., Deng, W., Wang, J., & Wang, S. Mechanism Analysis and Simulation Study of Automobile Millimeter Wave Radar Noise. SAE Technical Paper 2018-01-1641, 2018.